\documentclass[secnumarabic,aps,prl,showpacs,twocolumn,amsmath,amssymb,superscriptaddress,tightenlines]{revtex4-1}

\usepackage[colorlinks,citecolor=blue,linkcolor=cyan]{hyperref}
\usepackage[usenames,dvipsnames,svgnames,table]{xcolor}
\usepackage{grffile}
\usepackage[caption=false]{subfig}
\usepackage{graphicx}
\usepackage{mathtools}

\usepackage{mathptmx}
\usepackage{txfonts}

\newcommand{\sgn}{\textsf{sgn}}
\newcommand{\dg}{^{\dagger}}
\newcommand{\Tr}{\textsf{Tr}}
\newcommand{\pr}{^{\prime}}
\newcommand{\Eff}{\textsf{Eff}}

\newcommand{\TO}{\mathcal{T}}
\newcommand{\op}{\hat}
\newcommand{\ii}{\textsf{int}}
\newcommand{\Det}{\textsf{Det}}

\begin{document}
\date{\today}
\title{Odd frequency pairing of interacting Majorana fermions}

\author{Zhoushen Huang}
\affiliation{Institute for Materials Science, Los Alamos National Laboratory, NM 87545, USA}
\author{P.~W\"olfle}
\affiliation{Institut f\"ur Theorie der Kondensierten Materie,
Karlsruher Institut f\"ur Technologie, D-76131 Karlsruhe, Germany}
\affiliation{Institut f\"ur Nanotechnologie, Karlsruher Institut f\"ur Technologie, D-76021 Karlsruhe, Germany}
\author{A.~V.~Balatsky}
\email{avb@nordita.org}
\affiliation{Institute for Materials Science, Los Alamos National Laboratory, NM 87545, USA}
\affiliation{NORDITA, Roslagstullsbacken 23, SE-106 91\ \ Stockholm, Sweden}

\begin{abstract}
  Majorana fermions are rising as a promising key component in quantum
  computation. While the prevalent approach is to use a quadratic
  (i.e. non-interacting) Majorana Hamiltonian, when expressed in terms
  of Dirac fermions, generically the Hamiltonian involves interaction
  terms. Here we focus on the possible pair correlations in a simple
  model system.  We study a model of Majorana fermions coupled to a
  boson mode and show that the anomalous correlator between
  \emph{different} Majorana fermions, located at opposite ends of a
  topological wire, exhibits odd frequency behavior. It is stabilized
  when the coupling strength $g$ is above a critical value $g_c$.  We
  use both, conventional diagrammatic theory and a functional integral
  approach, to derive the gap equation, the critical temperature, the
  gap function, the critical coupling, and a Ginzburg-Landau theory
  allowing to discuss a possible subleading admixture of
  even-frequency pairing.
\end{abstract}
\pacs{}
\maketitle

\paragraph{Introduction} 

Recent advances in material fabrication techniques have brought
Majorana fermions out of a theoretician's toy box and placed them onto
the stage of realistic objects. Majorana fermions are particles with
the peculiar property of being their own anti-particles. While its
status as a fundamental particle is still unclear in the arena it was
first proposed, namely neutrino in particle physics, in condensed
matter physics, on the other hand, there have been realistic proposals
and indeed fairly convincing experimental signatures that these will
come into existence as zero energy
modes \cite{Kitaev,Alicea,DasSarma1,LR,Wilczek,LeoK,Marcus,wang14}. Apart from
the obvious theoretical interest, such Majorana zero modes are
particularly attractive as a potential key component in topological
quantum computation \cite{kitaev-qc03,nayak-rmp08,beenakker-arcmp13} because their origin is often of a topological
nature (vortices, end modes, topological Dirac materials, etc.), hence their
existence is robust and protected against disruptions such as defects,
perturbations and decoherence. Since they also exhibit anyonic
statistics under suitable conditions, their mutual braiding would
constitute the fundamental operation to encode and process
information. They are thus well suited for fault-tolerant quantum
computation.

While Majorana fermions are most often studied using an effective
Hamiltonian bilinear in Majorana operators, the fundamental object in
the condensed matter context, however, is still Dirac fermions. These
effective Hamiltonians, when written in terms of Dirac fermions,
generically contain both particle ($c\dg c$) and pairing ($cc$) terms,
which is why any condensed matter platform for Majorana fermions
necessarily contains superconductivity \cite{Wilczek}. As pairing
inevitably arises from interaction, an important question one should
ask is what types of instability would ensue. For conventional
fermions and bosons, such instabilities would generate non-vanishing
anomalous correlation functions which represent the onset of steady
orders such as charge, spin, or pair susceptibilities. The same can be
done for Majorana fermions \cite{Tanaka13,DasSarma15}. The effective Hamiltonian formalism, being
time-independent in its nature, implicitly assumes that these
instabilities are dominated by their equal time behavior (i.e. the
even-in-time component). However, the very nature of Majorana fermions
being simultaneously a particle and an anti-particle hints to the
possibility that time (or frequency) dependence in the order parameter
has a fundamental role to play. As a heuristic example, the pairing
correlator $M_{\tau} \equiv \langle \TO \mu(\tau) \mu(0)\rangle$ of
Majorana operator $\mu$ separated by (Matsubara) time $\tau$ is forced
to be \emph{odd in time}, and therefore vanishes at equal time. A more
complete discussion on the pairing symmetries of Majorana fermions
will be presented later.

Historically, an odd frequency (odd-f) pairing state was first pointed
out by Berezinskii \cite{Ber} in 1974 as a candidate state for
superfluid He3. While this was eventually proven not to be the case,
Berezinskii's pioneering paper set the stage for subsequent searches
of other possible odd-f states, such as spin singlet odd-f pairing
state \cite{BA,BK}. Although initial search of models and systems
realizing such states turned out to be inconclusive \cite{Abrahams},
more recent discussions of odd-f states have moved to systems of
superconducting heterostructures where one can have conventional pairs
converted into odd-f states at the interfaces
\cite{EfetovRMP,Tanakareview,Burset15,Ebisu15}. Moreover, odd-f states can also be
realized in homogeneous multiband superconductors \cite{BSBr,BSB}. If
we take the broader view that odd-f states represent a novel class of
hidden order, then this concept may be extended beyond
superconductivity to encompass a whole set of states that might have
odd-f order parameters \cite{Abrahams,EfetovRMP,Tanakareview}, e.g.,
spin nematics \cite{Nematicodd}, BEC \cite{bec}, and density waves
\cite{kedemodd}.

In this work, we will study pairing of Majorana fermions in the
presence of interaction induced by coupling to an external boson. Our
focus will be on the pairing between \emph{different} Majorana modes
because it arises solely as a consequence of interaction. Unlike
same-mode pairing which, as discussed before, is required to be odd-f
due to fermionic statistics, there is no \emph{a priori} requirement
on the frequency dependence for cross-mode Majorana pairing. Indeed,
as we will show, both even-f and odd-f solutions exist for the gap
equation, yet the odd-f solution has a lower free energy and is hence
more stable than the even-f state. We determine the phase diagram in
the temperature-coupling constant plane showing the disordered high
temperature phase and a pair-correlated low-temperature phase,
separated by a second-order transition. Interestingly, we find a
quantum phase transition at a critical coupling, in contrast to
conventional BCS theory where any weak attractive interaction induces
pairing, if only at an exponentially small temperature.  We derive the
gap equation and solve it both, analytically in the limit of high and
low temperature, and numerically. The critical coupling at low $T$ may
be linked to the first excited state of an effective Schr\"odinger
equation, enabling a determination of the critical coupling. We also
derive a functional integral representation of the partition function
in terms of the pair order field, which allows to estimate the
condensation energy. The latter may also be used to derive an
approximate Ginzburg-Landau expansion in terms of both, the odd- and
even-frequency order parameters.
 
\paragraph{Majorana pairing state and Extension of Berezinskii's
  classification}
To set the stage, first consider the notion of pairing state. A state
is paired when any two-particle anomalous expectation is finite: for
any particle operator $O^a(r,\tau)$, where $a$ denotes a collective
quantum number, its pairing amplitude is
 \begin{eqnarray}
P_O^{ab}(r,\tau) =  \langle \TO O^a(r,\tau)O^b(0,0)\rangle
\label{Pair1}
\end{eqnarray}
The cases of fermions $O(r,\tau) = c(r,\tau)$ , spins
$O^a(r,\tau) = S^a(r,\tau)$ ($a$ denoting directions), and boson
$O^a(r,\tau) = b_a(r,\tau)$ ($a$ denoting, say, species) all fall
within the category of thus defined paired states. This does not
necessarily mean there is a true superconducting or superfluid
order. For example, a typical pairing state that supports a superflow
would exhibit a $U(1)$ gauge symmetry breaking related to phase
coherence and off-diagonal long range order, and has other attributes
like phase stiffness. We instead accept that thus defined pairing
correlation is the basic property that we investigate in the context
of Majorana fermions. 
Since Majorana operators are real, the only symmetry left to be broken
is $Z_2$ symmetry. Thus there are important distinctions between the
``paired'' Majorana states and conventional pairing. Yet the question
about the appearance of anomalous order in models with Majorana states
stands and needs to be addressed. We focus on the pairing correlator
as a fundamental ingredient that any ``super'' state must possess --
such correlations might be important indicators of incipient orders.

\begin{figure}
  \centering
      \includegraphics[trim={3cm 2.5cm 3cm 2.5cm},width=0.35\textwidth,clip]{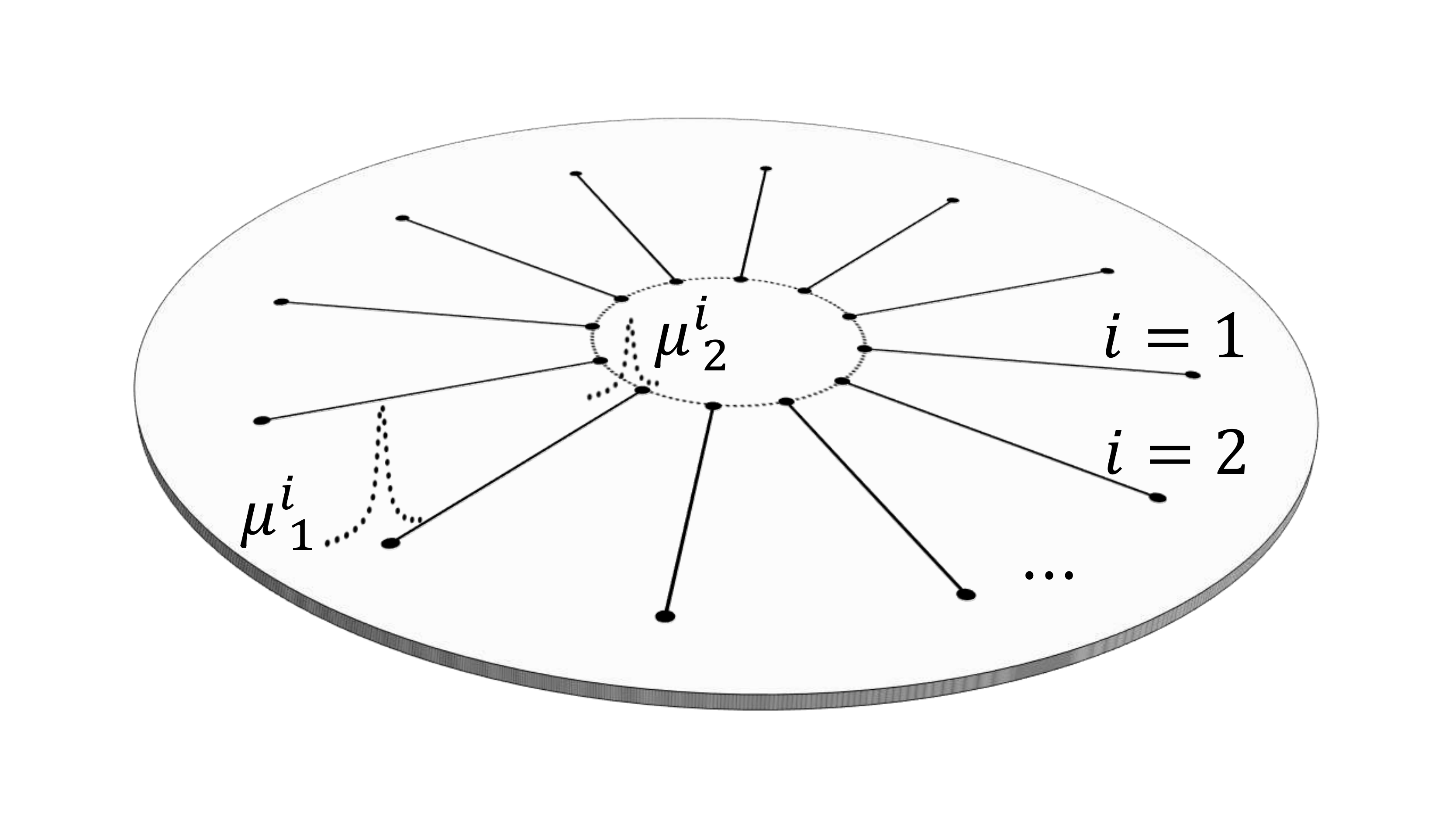}
      \caption{Schematics of superconducting wires that host Majorana
        fermions $\mu^i_a$ on their edges, where $i$ is the wire index
        and $a = 1,2$ labels the edge. Dashed line illustrates
        possible coupling between Majorana fermions on neighboring
        wires. }
  \label{fig:wire-ray}
\end{figure}

We start with the extension of the Berezinskii classification for
Majorana operators.  Consider a set of Majorana states that emerge
from materials like the ends of magnetic or semiconducting wires, see
Fig.~\ref{fig:wire-ray}. We introduce Majorana operators
 \begin{gather}
\mu_a^i = (\mu_a^i)^{\dag}\quad , \quad i   = 1...N\ , \ a = 1,2\ ,
\end{gather}
where $i$ is the wire index and $a$ denotes the two ends of the
wire. Following Eq.~\ref{Pair1}, consider the matrix of pair
amplitudes
\begin{gather}
  M_{ab}^{ij}(\tau) = \langle \TO\mu_a^i(\tau)\mu_b^j(0)\rangle =
  -M^{ji}_{ba} (-\tau)\ .
\label{Pair2}
\end{gather}
The antisymmetry follows from the definition of time ordering,
regardless of the Majorana nature of the operators and is similar to
Berezinskii's classification. Majorana fermion pairing is analogous to
same spin Dirac fermion pairing.

In the simplest case where $i=j$ and in the absence of any interaction, one has
 \begin{eqnarray}
M^{ii}_{ab}(\tau) \equiv M_{ab}(\tau) =M(\tau)\delta_{ab}= -M(-\tau)\delta_{ab}\ .
\label{Pair3}
\end{eqnarray}
Thus the {\em only pairing channel open} to a single Majorana fermion
is the odd-f pairing. Mixed pairs may form in the odd-f channel, with
a possible admixture of even-f component. From the free action
$S = \int d\tau \mu\partial_{\tau}\mu = 2
\sum_{\omega_n>0}i\omega_n\mu_{\omega_n}\mu_{-\omega_n}$,
the propagator, which is also the pairing amplitude, reads
\begin{gather}
  \label{m-free}
  M(\omega_n) = \frac{1}{2i\omega_n} \quad , \quad M(\tau) =
  \frac{1}{4}\sgn(\tau)\ .
\end{gather}
Thus a free Majorana state has odd-f pairing built in. Upon analytic
continuation, $M(\omega +i\delta) = M'(\omega) + i M''(\omega)$, with
\begin{gather}
  M'(\omega) = \textsf{Pf} \frac{1}{2\omega}\quad , \quad M''(\omega) =
 \frac{ \pi}{2} \delta(\omega)\ ,
\label{M2}
\end{gather}
where $\textsf{Pf}$ stands for principal factor.  Some comments are in
order: i) this analysis suggests that a free Majorana state at zero
energy is the simplest realization of odd-f state. Since the Majorana
fermion is both a particle and a hole, the single particle propagator
is a pair propagator. The natural question would be to identify the
pairing states that would emerge from interacting fermions, which we
shall address later. There are attempts to build in this direction
\cite{MF,Katsura}. ii) Real part of the pairing susceptibility is {\em
  odd} in frequency and time while the imaginary part is even. This
reversal of parities of real and imaginary parts with respect to the
case of BCS (even frequency pairing) is expected \cite{BA}. The
imaginary part $M''(\omega)$ that represents the decay of the pairs
has a sharp peak at zero energy, which indicates the strong decay of
odd-f pairing at zero frequency. i.e. at long times. The situation
here is similar but not identical to the case of odd-f states found in
superconducting heterostructures where a peak in DOS and in
$M''(\omega)$ is present and is associated with Andreev scattering
\cite{Tanakareview}.

While the analysis above is for zero-energy Majorana modes, the
inherent connection between single particle propagator and pairing
remains for the general case of a propagating Majorana branch of
exictations,
\begin{gather}
  G(k,\tau) = M(k,\tau) = \langle \TO \mu_{-k}(\tau)\mu_k(0)\rangle\ .
\end{gather}
In the remainder of this paper, therefore, we will not concern
ourselves with the effect of dispersion, but instead focus on how
interaction affects the fate of odd-f Majorana pairing.

\paragraph{Model of interacting Majorana fermions}
The minimal model for investigating the effect of interaction is to
consider a pair of Majorana fermions.
We focus in the following on a
single topological wire with particle-hole symmetric edge states.  Let
$c,c\dg$ be the electron operators, then the corresponding Majorana
operators $\mu_a, a=1,2$, are defined as coherent superpositions of
particles and
holes that satisfy
\begin{gather}
  c = \mu_1 + i\mu_2\quad , \quad c\dg = \mu_1 - i\mu_2\ .
\end{gather}
We note that $\{\mu_a(\tau),\mu_b(\tau)\}=\frac{1}{2}\delta_{ab}$. The
electron pair operator may be expressed as
$c(\tau)c(0)=\mu_1(\tau)\mu_1(0)-\mu_2(\tau)\mu_2(0)+
i\mu_1(\tau)\mu_2(0) + i\mu_2(\tau)\mu_1(0)$.
The point to note now is that whereas pair correlations of like
Majorana fermions are correlations in time of a single Majorana
fermion at either end of the wire, pair correlations of unlike
fermions, i.e. a finite amplitude of $\mu_1\mu_2$, describe nonlocal
correlations between Majorana fermions at opposite ends of the
wire. The latter amplitude is zero for free fermions, but will be
generated by interaction, as we will show below.

We now couple the Majorana fermions to a single boson mode of
frequency $\Omega$, as described by the Hamiltonian
\begin{gather}
  \label{hfull}
  H_{eb} =g (b\dg + b)i(\mu_1\mu_2-\mu_2\mu_1) +
  \Omega \Bigl(b\dg b + \frac{1}{2}\Bigr)\ ,
\end{gather}
where $g$ is the coupling strength. Such a boson mode can arise from
an optical phonon, for example. We have chosen to couple the boson
equally to both particle and hole, as
$i(\mu_1\mu_2 - \mu_2\mu_1) = c\dg c - \frac{1}{2} = \frac{1}{2}c\dg c
- \frac{1}{2} c c\dg$.
The Euclidean action corresponding to Eq.~\ref{hfull} is
$S_{eb} = \int d\tau\, \bar \psi \partial_{\tau} \psi + \bar \phi
(\partial_{\tau} + \Omega) \phi + g(\bar \phi + \phi)\bar \psi\psi$,
where $\psi$ and $\phi$ are fields for $c$ and $b$, respectively.
Integrating out the boson then leads to an interacting action of the
fermion,
$S \equiv -\log \Tr_{\phi} e^{-S_{eb}} = \int d\tau \,\bar
\psi \partial_{\tau}\psi - g^2 \int d\tau d\tau\pr\, \bar
\psi_{\tau}\psi_{\tau} K_{\tau,\tau\pr} \bar
\psi_{\tau\pr}\psi_{\tau\pr}$,
where
$K_{\tau,\tau\pr}= \frac{1}{2} \langle \TO
(b+b\dg)(\tau)(b+b\dg)(\tau\pr)\rangle$
is the boson propagator in time domain,
\begin{gather}
  \label{ktau} K_{\tau,\tau\pr} = \frac{1}{2}\left[ e^{-\Omega
|\tau\pr - \tau|} + \frac{2\cosh\left[\Omega (\tau\pr -
\tau)\right]}{e^{\beta \Omega} - 1} \right]\ ,
\end{gather}
and $\beta=T^{-1}$ is the inverse temperature. In the Majorana basis,
using the notation $\mu=\mu_1,\eta=\mu_2$, the effective action is
\begin{align}
  \label{s-maj}
  S = \int d\tau (\mu \partial_{\tau}\mu + \eta \partial_{\tau}\eta) 
  - 8 g^2 \int\limits_{\mathclap{\tau < \tau\pr}} d\tau d\tau\pr K_{\tau,\tau\pr} \mu_{\tau}\mu_{\tau\pr}\eta_{\tau}\eta_{\tau\pr}\ .
\end{align}
Since $K_{\tau,\tau\pr} = K_{\tau\pr,\tau}$, the $\tau > \tau\pr$
interaction is identical to that of $\tau<\tau\pr$ and is accounted
for by doubling the prefactor.

\paragraph{Gap equation for $\mu\eta$ pairing}
We now sketch out derivation of the gap equation. To decouple the
four-Majorana interaction, we employ Hubbard Stratonovich (HS)
transformation, after which the interaction term becomes (cf.~SM)
\begin{gather}
  \label{s-int-hs}
  S_{\textsf{int}}^{\textsf{HS}} = \int d\tau d\tau\pr \left[
    \frac{\Delta_{\tau,\tau\pr}\Delta_{\tau\pr,\tau}}{16g^2
      K_{\tau,\tau\pr}} - \Delta_{\tau,\tau\pr}\mu_{\tau}\eta_{\tau\pr}
  \right]\ ,
\end{gather}
where $\Delta_{\tau,\tau\pr}$ are complex HS fields subject to the
constraint
\begin{gather}
  \label{delta-constraint}
  \Delta_{\tau,\tau\pr} = \Delta_{\tau\pr,\tau}^{*}\implies
  \Delta_{\omega_n} \in \mathbb{R}\ ,
\end{gather}
which originates from the aforementioned redundancy of the
$\tau > \tau\pr$ interaction (cf.~SM), and the frequency domain
constraint follows assuming $\Delta$ depends only on $\tau\pr - \tau$.
\footnote{$\Delta_{\omega_n}$ can be understood as the statistical weight of the Majorana pair $\mu_{\omega_n}\eta_{-\omega_n}$.}
Integrating out $\mu$ and $\eta$, one obtains the effective action of
$\Delta$,
\begin{gather}
  \label{s-eff}
  S_{\Eff} = \sum_{\mathclap{\omega_m,\omega_n}}
  \frac{\Delta_{\omega_m}\Delta_{\omega_n}}{16g^2\beta}
  \left[\frac{1}{K}\right]_{\omega_m - \omega_n} \mathclap{-}
  \sum_{\omega_n>0} \log\left( \omega_n^2 -
    \frac{\Delta_{\omega_n}\Delta_{-\omega_n}}{4}\right)\ ,
\end{gather}
where $\omega_n = (2n+1)\pi T$ are fermionic Matsubara frequencies.
Demanding $\delta S_{\Eff}/\delta \Delta_{\omega_n} = 0$ then yields
the gap equation,
\begin{gather}
  \label{gap-eq}
  \frac{T}{2g^2} \sum_{\omega_m} \left[ \frac{1}{K} \right]_{\quad\ \mathclap{\omega_n - \omega_m}} \Delta_{\omega_m} = \frac{\Delta_{-\omega_n}}{\frac{1}{4} \Delta_{\omega_n}\Delta_{-\omega_n} - \omega_n^2}\ .
\end{gather}
This equation may be expressed in the more familiar form
\begin{gather}
  \label{gap-eq2}
\Delta_{\omega_n}=8g^2T \sum_{\omega_m}K_{{\omega_n - \omega_m}}M^{21}_{\omega_m}.
\end{gather}
where
$M^{21}=(\Delta^{-}-\Delta^{+})/[4\omega_n^2+
(\Delta^{-})^2-(\Delta^{+})^2]$
is the anomalous Green's function (cf.~Eq.\ref{Pair2}) found by
inverting the matrix
\begin{gather}
  M^{-1}=2i\omega_n-\sigma^{x}\Delta^{-}_{\omega_n}-i\sigma^{y}\Delta^{+}_{\omega_n}.
\end{gather}
where
$\Delta^{\pm}_{\omega_n} = \frac{1}{2}(\Delta_{\omega_n} \pm
\Delta_{-\omega_n})$,
and the phonon propagator
$K_{\nu_n}= \Omega/[(\Omega)^2 - (i\nu_n)^2]$ is the Fourier transform
of Eq.~\ref{ktau}. Here we used the symmetry property of
Eq.~\ref{Pair2}. It is worth mentioning that the gap equation
Eq.~\ref{gap-eq2} differs from the conventional gap equation by a
factor of $4$.  This is due to the fact that the diagram rules for
Majorana fermions as compared to usual fermions have to be modified by
applying a factor of $n$ at each interaction vertex involving $n$
Majoranas (at the electron-phonon vertex this would be a factor of 2),
see also SM.

\paragraph{Emergence of odd-f pairing} We now discuss the solutions
to the gap equation.  The normal state ($\Delta_{\omega_n} = 0$) is a
trivial solution.  Nonzero solutions can be found analytically at high
and low temperatures. At high temperatures $T>>\Omega$, the kernel may
be taken to be diagonal, 
$K_{\omega_n - \omega_m} = \delta_{mn} /\Omega = \delta_{mn}
K_{\tau,\tau}/T$
where $K_{\tau,\tau}$ is the equal time boson propagator, see Eq.~\ref{ktau}.
Eq.~\ref{gap-eq2} only has pure odd-f and even-f solutions,
\begin{gather}
  \label{gap-eq-sol-0}
  \Delta_{\omega_n}^{(s)} = s\Delta_{-\omega_n}^{(s)} = 2
  \sqrt{2g^2 K_{\tau,\tau} +s \omega_n^2}\quad , \quad s = \pm 1\ .
\end{gather}
We emphasize that no assumption was made on how $\Delta$ transforms
under $\omega_n \leftrightarrow -\omega_n$, yet the saddle point
solutions acquire a symmetry. The pure odd-f (even-f) solution is
found to have a lower (higher) free energy than the unpaired state
(cf.~SM), hence we will focus on the odd-f case. 
Stability of the odd-f state is consistent with Ref.~\cite{SMM09} and is discussed in more detail in SM.
The reality constraint Eq.~\ref{delta-constraint} imposes a frequency
cutoff $\omega_c=\sqrt{2g^2T/\Omega}$ for the odd-f solution above
which the gap function falls back to zero.
Since the smallest possible $|\omega_n|$ is $\omega_0 = \pi T$,
requiring $\omega_c \ge \omega_0$ entails
that there is a critical temperature $T_c$ above which the system is
normal,
\begin{gather}
  \label{T-c}
  T_c = \frac{2 g^2}{\pi^2 \Omega}\ .
\end{gather}
Closely below $T_c$, the first nonzero gap component
$\Delta^{-}_{\omega_0}$ has the usual mean field temperature
dependence
\begin{gather}
\label{del}
\Delta^{-}_{\omega_0} =2\pi T_c \sqrt{1-T/T_c}\ .
\end{gather}

At low temperatures $T \ll \Omega$, the kernel $K_{\nu_n}$ is
approximately constant for $\nu_n < \Omega$ and drops as $\nu_n^{-2}$
beyond. Therefore we expect $\Delta_{\omega_n}$ to be maximum at
$\omega_n \sim \Omega$. Approximating $\Delta$ by its value at this
typical frequency on both sides of Eq.~\ref{gap-eq2}, one is left with
the frequency summation
$T\sum_{\nu_n}K_{\nu_n} = K_{\tau,\tau} \sim \frac{1}{2}$. The
resulting $\Delta_{\omega_n = \Omega}$ is again given by
Eq.~\ref{gap-eq-sol-0}, hence a real valued solution can only be found
for $g>g_c \sim \Omega$.

\begin{figure}
  \centering
    \includegraphics[width=0.45\textwidth]{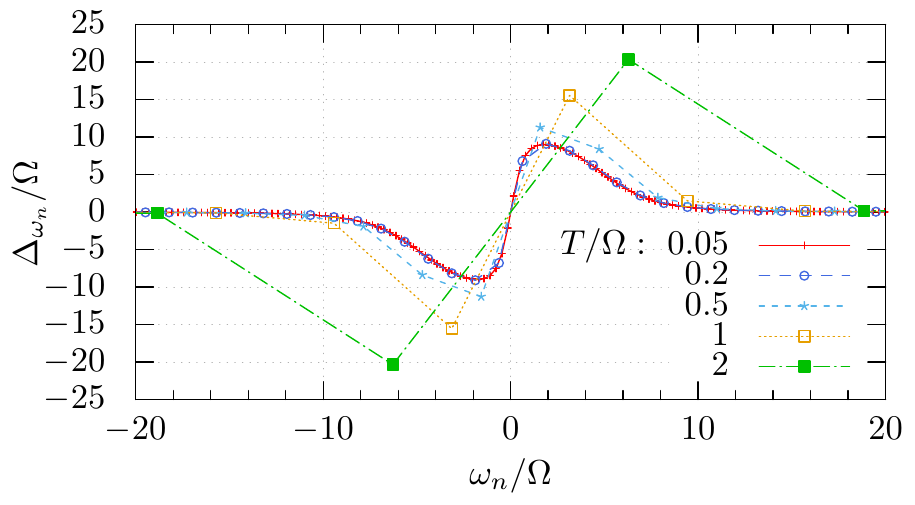}
    \caption{(Color online) Pairing of different Majorana fermions.
      Coupling strength is set to $g = 6\Omega$. The gap function
      peaks at the larger of the two energy scales, temperature
      $\pi T$ or boson frequency $\Omega$. }
    \label{fig:delta}
\end{figure}

For generic temperatures, we have obtained the gap function
numerically by minimizing the action
Eq.~\ref{s-eff}. Fig.~\ref{fig:delta} shows results of
$\frac{\Delta_{\omega_n}}{\Omega}$ versus $\frac{\omega_n}{\Omega}$
for a fixed value of $\frac{g}{\Omega}=6$ and several values of
$T<T_c$. The peak sits at the dominant energy scale:
$\omega_{\textsf{peak}} = \pi T$ for $T \gg \Omega$, and
$\omega_{\textsf{peak}} \sim \Omega$ for $T \rightarrow 0$.  The slope
$\frac{\Delta_{\omega_{\textsf{peak}}}}{\omega_{\textsf{peak}}}$
serves as an order parameter for odd-f pairing.

\begin{figure}
  \centering
    \includegraphics[width=0.44\textwidth]{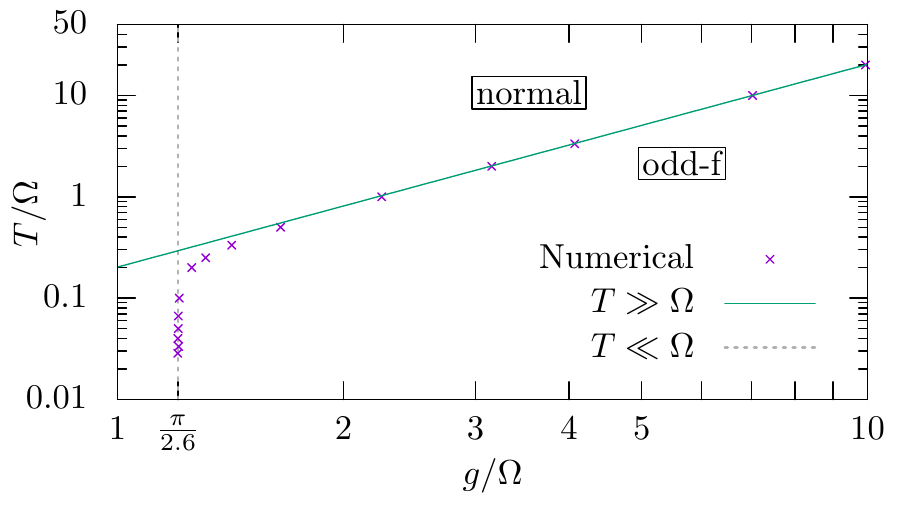}
    \caption{(Color online) Phase diagram in log scale at
      $\Omega = 1$. Colored crosses: numerical results for the phase
      boundary $T_c(g)$. Green solid line: high $T$ limit
      Eq.~\ref{T-c}. Gray dashed line: $T\rightarrow 0$ limit
      Eq.~\ref{g-c-0}.}
    \label{fig:phase-diag}
\end{figure}

\paragraph{Critical coupling and effective Schr\"odinger
  equation}
The $T \ll \Omega$ analysis above indicates the possibility of a
critical coupling $g_c \sim \Omega$ at zero temperature. This agrees
with our numerically determined phase diagram
Fig.~\ref{fig:phase-diag}.
To clarify this question, 
we now analyze the gap equation at $T = 0$ and $\Delta \rightarrow 0$,
which may then be linearized and transformed into a differential
equation in time
\begin{gather}
  \label{schroed-eq}
\left[ -\frac{d^2}{dx^2} - V(x)\right] Y(x)=0\ ,
\end{gather}
where $x = \tau\pr - \tau$, $Y(x)=-Y(-x)$ is the Fourier transform of
$\Delta^-_{\omega}/\omega^2$, and $V(x) = g^2 e^{-\Omega |x|}$. This
is a ``Schr\"odinger equation'' for energy eigenvalue $E=0$. The
condition for the lowest-lying \emph{odd-parity} state to have energy
$E=0^{-}$
is found as
\begin{gather}
  \label{g-c-0}
  g_c(T=0) = \frac{\pi}{\gamma} \Omega\ ,
\end{gather}
where $\gamma = 2$ if $V(x)$ is approximated by a square well of depth
$g^2$ and width $\Omega^{-1}$, in which case $Y(x)\propto \sin(g_c x)$
inside the potential well and decays rapidly on the outside, hence
$\Delta_{\omega_n}^{-}$ peaks at $\omega_{\textsf{peak}} = g_c$.
Numerically we found $\gamma \simeq 2.6$, see
Fig.~\ref{fig:phase-diag}. We have a quantum critical point at $g=g_c$
separating the normal phase from the paired phase.

\paragraph{Ginzburg-Landau theory} In order to understand the
interplay of odd-f and even-f pairing better we now formulate a
Ginzburg-Landau free energy expression. As a suitable order parameter
for odd-f pairing we choose the slope $A$ of the frequency dependent
gap function, $\Delta^{-}=A \omega$ in the low frequency regime
$|\omega_n|<\omega_x$. The even-f order parameter we define as
$B=\Delta^{+}/\Omega$.  Then, the free energy of low frequency modes
as derived from Eq.\ref{s-eff} may be approximated as (see SM)
 \begin{gather}
F=c_0\Bigl[\left(t-1\right) A^2+ \frac{A^4}{8}\Bigr]
+\Bigl[\left(b_0 t
  +4c_{1}\right)B^2+ 2 c_{2} B^4\Bigr] -  c_{1} A^2 B^2
  \label{G-L}
\end{gather}
where $t \equiv \frac{T}{T_c}$,
$b_{\ell} = T_c \sum_{m,n}\pr\frac{( \omega_m
  \omega_n)^{\ell}}{16g^2 \Omega^{2(\ell-1)}}[K^{-1}]_{mn}$,
$c_\ell= \frac{1}{4} \sum_{n}\pr[\frac{\Omega}{2\omega_n}]^{2\ell}$ and
$\sum\pr$ is restricted to $|\omega_n|<\omega_x$, with $\omega_x$
chosen such that $b_1=c_0$. It is seen that as found numerically odd-f
pairing appears for $T<T_c$ (provided $g>g_c$), whereas even-f pairing
always increases the free energy, even taking into account the
negative contribution of the mixed term (for details, see SM).

\paragraph{Conclusion}
We extended Berezinskii's classification of pairing order parameters to Majorana
fermions, and showed that odd frequency pairing arises naturally in
both free and interacting Majorana theories. For free Majoranas, odd-f
pairing is the only pairing channel allowed by the fermionic exchange
statistics. With interaction, pairing of Majorana fermions at opposite ends of a topological wire 
may be induced. We show that within a model of phonon-mediated 
interaction there is a critical coupling strength for pairing, which implies the existence 
of a quantum critical point, and can be determined by mapping 
the gap equation onto an effective Schr\"{o}dinger equation.  We find possible 
pairing solutions both analytically and numerically and show that only odd-f pairing is stable, 
although even-f pairing is in principle allowed. To this end we define two frequency-independent order parameters characterizing both odd-f and even-f pairing and propose an effective 
Ginzburg-Landau free energy derived from the microscopic mean field action. 
The simple model study of Majorana pairing presented here opens the way for a study of more complex systems involving many Majorana states, e.g. located at opposite edges of a two-dimensional topological superconductor.

\paragraph{Note added} We recently became aware of a related work by
Zhang and Nori \cite{Zhang15}.

\paragraph{Acknowledgments}
We are grateful to A.~Black-Schaffer, P.~Brower, F.~von Oppen,
H.~Katsura, Y.~Kedem, K.~Zarembo, and D.~P.~Arovas for useful
discussions.  This work was supported by the US DOE Basic Sciences for
the National Nuclear Security Administration of the US Department of
Energy under Contract No. DE-AC52-06NA25396, E304, the Knut and Alice
Wallenberg Foundation, and the European Research Council under the
European Union's Seventh Framework Program (FP/2207-2013)/ERC Grant
Agreement No. DM-321031.

\bibliography{ref}
\bibliographystyle{apsrev-no-url}

\onecolumngrid
\appendix*
\section{Supplementary Material}
\setcounter{equation}{0}
In this note, we derive the gap equation in detail using both field
theory and diagrammatic expansion. Then we discuss the free energy of
the odd-f and even-f pairing fields.
\section{Field theoretic derivation of the gap equation}
The Majorana action from the text is
\begin{gather}
  \label{sm-s-maj}
  S = S_0 + S_{\ii}\quad , \quad 
  S_0 = \int d\tau (\mu \partial_{\tau}\mu + \eta \partial_{\tau}\eta) \quad , \quad
  S_{\ii} = 
  - 8 g^2 \int\limits_{\mathclap{\tau < \tau\pr}} d\tau d\tau\pr K_{\tau,\tau\pr} \mu_{\tau}\mu_{\tau\pr}\eta_{\tau}\eta_{\tau\pr}\ ,\\
  \label{sm-ktau} K_{\tau,\tau\pr} = \frac{1}{2}\left[ e^{-\Omega
      |\tau\pr - \tau|} + \frac{2\cosh\left[\Omega (\tau\pr -
        \tau)\right]}{e^{\beta \Omega} - 1} \right]\quad , \quad
  K_{\nu_n} = \frac{\Omega}{\Omega^2 + \nu_n^2}\ .
\end{gather}
Since $K_{\tau,\tau\pr} = K_{\tau\pr,\tau}$, the $\tau > \tau\pr$
interaction is equivalent to that of $\tau < \tau\pr$ and is accounted
for by doubling the prefactor in $S_{\ii}$. The purpose of keeping
track of such redundancy in the interacting action is to avoid
introducing multiple Hubbard-Stratonovich (HS) fields for equivalent
interaction terms.

HS transformation uses the following identity: for a real positive
number $M$ and any four Grassmann variables
$\psi_1, \psi_2, \psi_3, \psi_4$, 
\begin{gather}
  \label{sm-useful-id}
  e^{M\psi_1\psi_2\psi_3\psi_4} = e^{M\psi_1\psi_2\psi_3\psi_4}
  \times \left[\int d\hat\Delta d\hat\Delta^{*}\, e^{-(\hat\Delta^{*} - M
      \psi_1\psi_2)M^{-1}(\hat\Delta - M \psi_3\psi_4)} \right]
  = \int d\hat\Delta d\hat\Delta^{*} \exp\left[- \frac{|\hat\Delta|^2}{M} +
    \hat\Delta \psi_1\psi_2 + \hat\Delta^{*}\psi_3\psi_4 \right]\
  ,
\end{gather}
upon proper normalization of the integration measure
$d\op \Delta d\op\Delta^{*}$. To decouple $S_{\ii}$ in the $\mu\eta$
pairing channel, we use Eq.~\ref{sm-useful-id} and introduce, for each
pair of time indices $\tau,\tau\pr$ with $\tau < \tau\pr$, a complex
field $\hat\Delta_{\tau,\tau\pr}$. After HS transformation, $S_{\ii}$
becomes
\begin{gather}
  \label{sm-maj-int-hs}
  S_{\textsf{int}}^{\textsf{HS}} = \int\limits_{\mathclap{\tau <
      \tau\pr}}d\tau d\tau\pr\,
  \left[\frac{\left|\hat\Delta_{\tau,\tau\pr}\right|^2}{8g^2
      K_{\tau,\tau\pr}} - \hat\Delta_{\tau,\tau\pr}
    \mu_{\tau}\eta_{\tau\pr} - \hat\Delta^{*}_{\tau,\tau\pr}
    \mu_{\tau\pr}\eta_{\tau} \right]\ .
\end{gather}
Note that $\hat\Delta$ is only defined for $\tau < \tau\pr$. One can
introduce a new field $\Delta_{\tau,\tau\pr}$ with unrestricted time
indices,
\begin{gather}
  \label{sm-del}
  \Delta_{\tau,\tau\pr} =
  \begin{cases}
    \hat\Delta_{\tau,\tau\pr} & \tau < \tau\pr \\
    \hat\Delta^{*}_{\tau\pr,\tau} & \tau > \tau\pr
  \end{cases}\ .
\end{gather}
If $\Delta_{\tau,\tau\pr}$ only depends on the time difference
$\tau\pr - \tau$, then Eq.~\ref{sm-del} entails that the Fourier
transform of $\Delta(\tau\pr - \tau)$ is real,
\begin{gather}
  \label{sm-del-real}
  \Delta_{\omega_n} = \int d\tau \Delta(\tau) e^{i\omega_n \tau} =
  \int\limits_{\tau > 0} d\tau \left(\Delta(\tau) e^{i\omega_n\tau} +
  \text{c.c.}\right) \in \mathbb{R}\ .
\end{gather}
Using the $\Delta$ fields, Eq.~\ref{sm-maj-int-hs} can be rewritten as
\begin{gather}
  \label{sm-s-int-hs}
  S_{\textsf{int}}^{\textsf{HS}} = \int d\tau d\tau\pr \left[
    \frac{\Delta_{\tau,\tau\pr}\Delta_{\tau\pr,\tau}}{16g^2
      K_{\tau,\tau\pr}} - \Delta_{\tau,\tau\pr}\mu_{\tau}\eta_{\tau\pr}
  \right]\ .
\end{gather}
Reinstating $S_0$ and going to frequency domain, the full action is
\begin{gather}
  \label{sm-s-hs}
  S^{\textsf{HS}} = \sum_{\omega_m,\omega_n}
  \frac{\Delta_{\omega_m}\Delta_{\omega_n}}{16g^2\beta} \left[
    \frac{1}{K}\right]_{\omega_n - \omega_m} + 
  \underbrace{\sum_{\omega_n > 0}
  (\mu_{\omega_n} \ , \ \eta_{\omega_n})
  \overbrace{\begin{pmatrix}
    2i\omega_n & -\Delta_{\omega_n} \\ \Delta_{-\omega_n} & 2i\omega_n
  \end{pmatrix}}^{\equiv \op G_{\omega_n}^{-1}}
  \begin{pmatrix}
    \mu_{-\omega_n} \\ \eta_{-\omega_n}
  \end{pmatrix}}_{\equiv S_{\textsf{Fer}}}\ .
\end{gather}
The fermion part is bilinear and can be integrated out,
\begin{gather}
  Z_{\textsf{Fer}} = \int D(\bar \psi,\psi) e^{-S_{\textsf{Fer}}} =
  \prod_{\omega_n > 0} \left[-\frac{1}{4} \Det \op G_{\omega_n}^{-1}
  \right] = \prod_{\omega_n > 0} (\omega_n^2 - \frac{1}{4}\Delta_{\omega_n}\Delta_{-\omega_n})\ ,
\end{gather}
where the factor $-1/4$ comes from the Jacobian
$\partial(\mu_{\omega},\mu_{-\omega},\eta_{\omega},\eta_{-\omega})/\partial(\bar
\psi_{\omega},\psi_{\omega},\bar\psi_{-\omega},\psi_{-\omega})$.
The effective action of $\Delta$ thus reads
\begin{gather}
  \label{sm-s-eff}
  S_{\Eff} = \sum_{\mathclap{\omega_m,\omega_n}}
  \frac{\Delta_{\omega_m}\Delta_{\omega_n}}{16g^2\beta}
  \left[\frac{1}{K}\right]_{\omega_m - \omega_n} \mathclap{-}
  \sum_{\omega_n>0} \log\left( \omega_n^2 -
    \frac{1}{4}\Delta_{\omega_n}\Delta_{-\omega_n}\right)\ ,
\end{gather}
and minimization with respect to $\Delta$ yields the gap equation
\begin{gather}
  \label{sm-gap-eq}
  \frac{1}{2g^2\beta} \sum_{\omega_m} \left[ \frac{1}{K}
  \right]_{\quad\ \mathclap{\omega_n - \omega_m}} \Delta_{\omega_m} =
  \frac{\Delta_{-\omega_n}}{\frac{1}{4}
    \Delta_{\omega_n}\Delta_{-\omega_n} - \omega_n^2}\ .
\end{gather}

\section{Gap equation from diagrammatic expansion}
In this section, we treat the interacting action $S_{\ii}$
(Eq.~\ref{sm-s-maj}) as a perturbation, and derive the gap equation as
the one-loop correction to the self energy. We will use $\nu$ and
$\omega$ to denote bosonic and fermionic Matsubara frequencies,
respectively.

\begin{figure}
  \centering
  \subfloat[][Interaction vertex]
  {
    \includegraphics[width=0.35\textwidth]{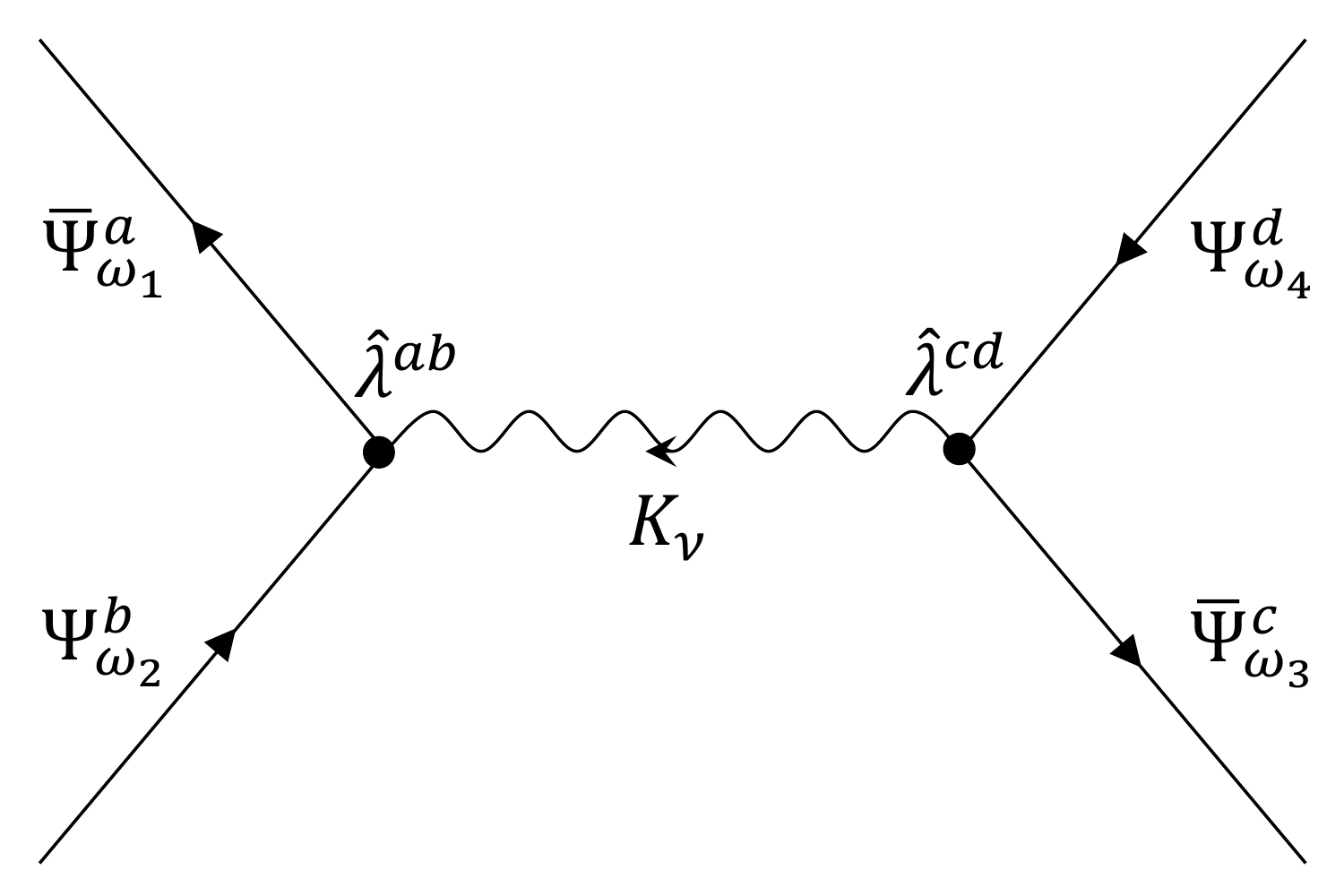}
    \label{sm-fig:vertex}
  }
  \subfloat[][Self energy]
  {
    \includegraphics[trim=0 1.5cm 0 2cm,width=0.5\textwidth,clip]{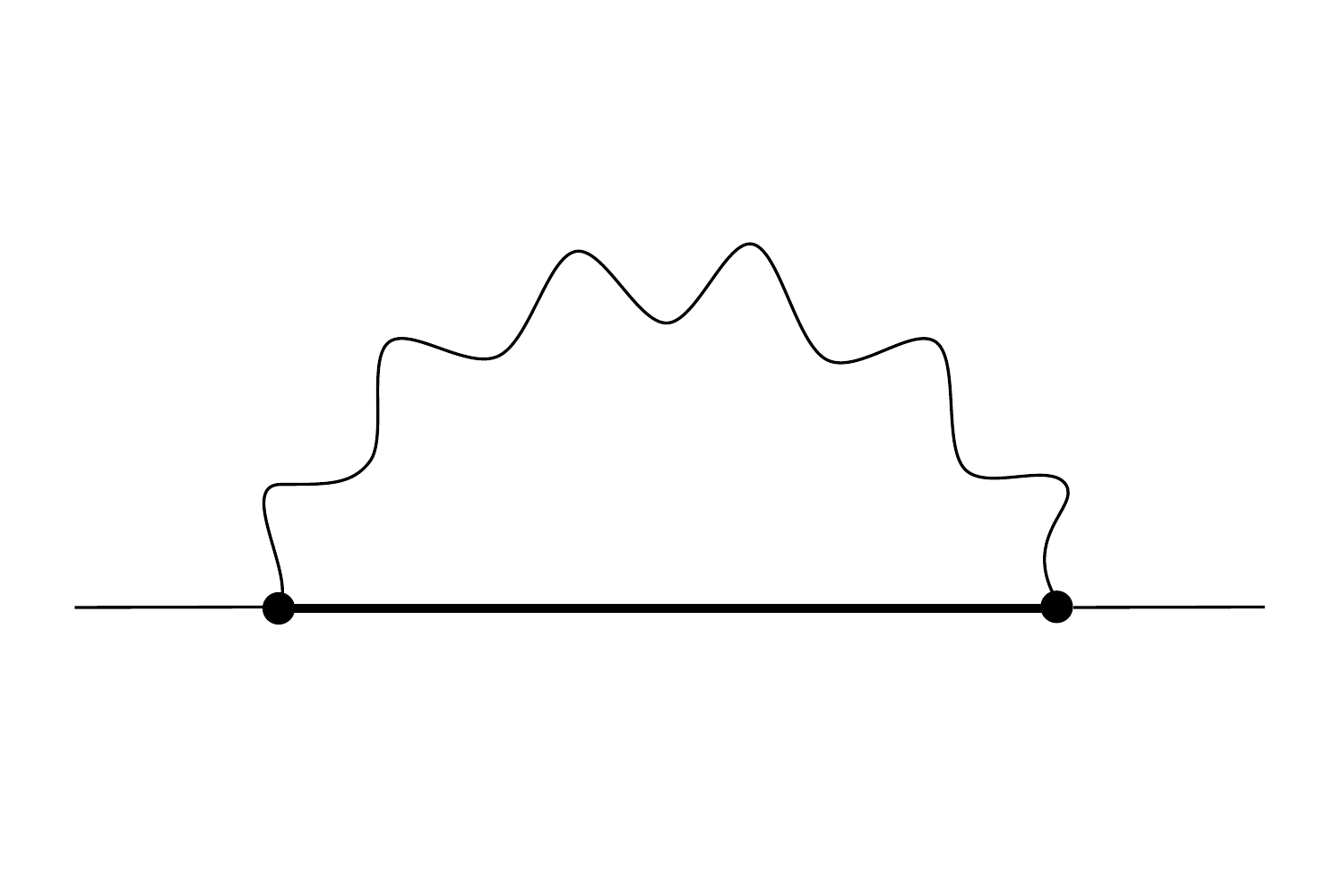}
    \label{sm-fig:self-erg}
  }
  \caption{Feynman diagrams. (a): Interaction vertex, see
    Eq.~\ref{sm-sint-vertex}. Frequencies going into the same
    $\op\lambda$ vertex sum to zero, and arrows denote the sign of the
    corresponding frequency in the $\delta$ function. (b): Self energy
    at the one loop level, see discussion below
    Eq.~\ref{sm-connected-contraction} for multiplicity. Thick line
    represents the dressed Green's function used in the self
    consistent equation Eq.~\ref{sm-sce}.}
  \label{sm-fig:feynman-diag}
\end{figure}

Let us introduce two-component Grassmann vectors,
\begin{gather}
  \Psi_{\omega} =
  \begin{pmatrix}
    \mu_{\omega}\\ \eta_{\omega}
  \end{pmatrix}\quad , \quad \bar\Psi_{\omega} = (\mu_{-\omega}\ , \ \eta_{-\omega})\ .
\end{gather}
The matrix Green's function is defined as
\begin{gather}
  \op G_{\omega} \equiv \langle \Psi_{\omega}\bar \Psi_{\omega}\rangle = \Bigl<
  \begin{pmatrix}
    \mu_{\omega}\mu_{-\omega} & \mu_{\omega}\eta_{-\omega} \\ \eta_{\omega}\mu_{-\omega} & \eta_{\omega}\eta_{-\omega}
  \end{pmatrix}\Bigr>\ .
\end{gather}
It is equivalent to the Majorana pairing matrix $M$ defined in the
text. One has
\begin{gather}
  \label{sm-fullG}
  \hat G_{\omega}^{\alpha\beta} = \frac{\langle \Psi_{\omega}^{\alpha} \bar
    \Psi_{\omega}^{\beta} e^{-S_{\ii}}\rangle_0}{\langle
    e^{-S_{\ii}}\rangle_0}
\end{gather}
where $\langle \cdots \rangle_0$ denotes averaging over the free
action $S_0$, and in the vector notation, the interacting action can
be written as
\begin{align}
  -S_{\ii} &= g^2 \int d\tau d\tau\pr K_{\tau\tau\pr} \left[\bar\Psi_{\tau} \op\lambda \Psi_{\tau} \right] \left[\bar\Psi_{\tau\pr} \op\lambda \Psi_{\tau\pr} \right]\\
  \label{sm-sint-vertex}
  &= g^2 T\sum_{\nu, \omega_{1,2,3,4}} K_{\nu} \left[\bar \Psi_{\omega_1} \op\lambda \Psi_{\omega_2} \right] \left[\bar \Psi_{\omega_3} \op\lambda \Psi_{\omega_4} \right] \delta(\omega_1 - \omega_2 - \nu)\delta(\omega_3-\omega_4+\nu)\ ,
\end{align}
where the matrix $\op \lambda = -\sigma_y$. The summand corresponds to
the diagram Fig.~\ref{sm-fig:vertex}.
To the order of $g^2$, Eq.~\ref{sm-fullG} becomes
\begin{gather}
  \label{sm-fullG-g2}
  \op G_{\omega}^{\alpha\beta} = \frac{\langle \Psi_{\omega}^{\alpha} \bar \Psi_{\omega}^{\beta} (1-S_{\ii})\rangle_0}{1-\langle S_{\ii}\rangle_0} = \langle\Psi_{\omega}^{\alpha}\bar\Psi_{\omega}^{\beta}\rangle_0 + \langle \Psi_{\omega}^{\alpha}\bar\Psi_{\omega}^{\beta} \times (-S_{\ii})\rangle_0^c + \mathcal{O}(g^4)\ ,
\end{gather}
where the superscript $c$ denotes connected contractions, i.e. both
$\Psi_{\omega}^{\alpha}$ and $\bar\Psi_{\omega}^{\beta}$ are to be
contracted with fields in $S_{\ii}$, not with each other. Explicitly
written out, one has (summing over repeated indices)
\begin{gather}
  \label{sm-connected-contraction}
  \langle \Psi_{\omega}^{\alpha}\bar\Psi_{\omega}^{\beta} \times
  (-S_{\ii})\rangle_0^c = g^2 T\sum_{\nu, \omega_{1,2,3,4}} K_{\nu}
  \delta(\omega_1 - \omega_2 - \nu)\delta(\omega_3-\omega_4+\nu) \hat\lambda^{ab}\hat\lambda^{cd} 
  \times
  \langle \Psi_{\omega}^{\alpha} \times \bar\Psi_{\omega_1}^a\Psi_{\omega_2}^b \bar\Psi_{\omega_3}^c\Psi_{\omega_4}^d \times \bar\Psi_{\omega}^{\beta}\rangle_0^c\ .
\end{gather}
If we ignore the Hartree terms in $\langle\cdots\rangle_0^c$, i.e.,
those with the two external fields $\Psi^{\alpha}$ and
$\bar \Psi^{\beta}$ contracting on the same $\hat\lambda$ vertex, then
it corresponds to the diagram Fig.~\ref{sm-fig:self-erg}, and since there
is no distinction between particle and hole for Majorana fermions,
$\Psi^{\alpha}$ can contract with any of $a,b,c,d$, and
$\bar\Psi^{\beta}$ can then contract with two of $a,b,c,d$ on the
opposite vertex of $\Psi^{\alpha}$. There are $4\times 2$ such
contractions of identical results. For example, consider the
contraction of
$\langle\alpha,b\rangle \langle a,c\rangle \langle d,\beta\rangle$.
The two ``anomalous'' ones are
\begin{gather}
  \langle\Psi^{\alpha}_{\omega} \Psi^b_{\omega_2}\rangle_0 = \delta_{\omega, -\omega_2} G_{\omega}^{(0)\alpha b}\quad , \quad
  \langle \bar\Psi_{\omega_1}^{a} \bar\Psi^c_{\omega_3}\rangle_0 = \delta_{\omega_1, - \omega_3} G_{-\omega_1}^{(0)a c}\ .
\end{gather}
From the delta functions one has $\omega_4 = -\omega_2 = \omega$ and
$\omega_3 = -\omega_1 = \omega - \nu$.  The contribution to the RHS
of Eq.~\ref{sm-connected-contraction} is thus
\begin{gather}
  \label{sm-bacd}
  - g^2 T\sum_{\nu} K_{\nu} \op\lambda^{ab}\op \lambda^{cd} \times \op
  G_{\omega}^{(0)\alpha b} \op G_{\omega - \nu}^{(0)ac} \op
  G_{\omega}^{(0)d\beta} = g^2T \left[
    \op G_{\omega}^{(0)}\op \lambda \left(\sum_{\nu} K_{\nu} \op G^{(0)}_{\omega - \nu}\right) \op\lambda \op G_{\omega}^{(0)}
  \right]^{\alpha\beta}\ ,
\end{gather}
where the minus sign comes from exchanging the order of $a$ and $b$ in
the contraction, and we used $-\op\lambda^T =
\op\lambda$. Eq.~\ref{sm-connected-contraction} is then
\begin{gather}
  \langle \Psi_{\omega}^{\alpha}\bar\Psi_{\omega}^{\beta} \times
  (-S_{\ii})\rangle_0^c = \left[ \op G_{\omega}^{(0)} \op
    \Sigma_{\omega} \op G_{\omega}^{(0)}\right]^{\alpha\beta}
\end{gather}
where
\begin{gather}
  \label{sm-self-erg}
  \op\Sigma_{\omega} = 8\times g^2 T\sum_{\nu} K_{\nu} \left[\op\lambda \op G_{\omega - \nu}^{(0)} \op \lambda\right]\ .
\end{gather}
Comparing Eq.~\ref{sm-fullG-g2} with the definition of self energy,
\begin{gather}
  \Sigma_{\omega} \equiv (G^{(0)}_{\omega})^{-1} - G_{\omega}^{-1}
  \Leftrightarrow G_{\omega} = G_{\omega}^{(0)} +
  G_{\omega}^{(0)}\Sigma_{\omega} G_{\omega}^{(0)} +
  G_{\omega}^{(0)}\Sigma_{\omega} G_{\omega}^{(0)}\Sigma_{\omega}
  G_{\omega}^{(0)} + \cdots\ ,
\end{gather}
one can identify that up to leading correction, Eq.~\ref{sm-self-erg} is
indeed the self energy. Replacing $\op G^{(0)}$ with the full $\op G$
on the RHS of Eq.~\ref{sm-self-erg} then leads to the self consistent
equation for the self energy matrix,
\begin{gather}
  \label{sm-sce}
  \op\Sigma_{\omega} = 8\times g^2 T\sum_{\nu} K_{\nu} \left[\op\lambda \op G_{\omega - \nu} \op \lambda\right]\ .
\end{gather}
Now writing
\begin{gather}
  \op G_{\omega}^{-1} \equiv
  \begin{pmatrix}
    2i\omega & -\Delta_{\omega} \\ \Delta_{-\omega} & 2i\omega
  \end{pmatrix}\implies \op G_{\omega} = \frac{
    \begin{pmatrix}
      2i\omega & \Delta_{\omega} \\ -\Delta_{-\omega} & 2i\omega
    \end{pmatrix}
}{\Det \op G^{-1}_{\omega}} \implies \op \lambda \op G_{\omega}\op \lambda = \frac{\begin{pmatrix}
  2i\omega & \Delta_{-\omega} \\ -\Delta_{\omega} & 2i\omega
\end{pmatrix}
}{\Det \op G_{\omega}^{-1}}\ ,
\end{gather}
note that $\op G^{-1}$ is the same fermion kernel as one would get in
the field theoretic approach, cf.~Eq.~\ref{sm-s-hs}. Then the self
consistent equation for the matrix element $\op \Sigma_{\omega}^{12}$
reads
\begin{gather}
  \label{sm-gap-eq-2}
  \Delta_{\omega} = 8g^2 T\sum_{\omega\pr} K_{\omega - \omega\pr}
  \frac{\Delta_{-\omega\pr}}{\Delta_{\omega\pr}\Delta_{-\omega\pr} -
    4{\omega^{\prime}}^2}\ ,
\end{gather}
which is the same as Eq.~\ref{sm-gap-eq} obtained from minimizing the
HS action. Eqs.~\ref{sm-gap-eq} and \ref{sm-gap-eq-2} appear to have
different factors of $T$, this is due to the convention of the Fourier
transform $A_{\nu} = \int d\tau A_{\tau} e^{i\nu\tau}$, and the fact
that $K_{\tau}$ is dimensionless, thus both $K_{\nu}$ and
$(1/K)_{\nu}$ have a dimension of time, hence the different factors of
$T$.

\section{Free energy consideration of the odd-f and even-f pairing
  states}
Consider the action Eq.~\ref{sm-s-eff} relative to the normal state
$\Delta = 0$ (which is always a solution of the gap equation),
\begin{gather}
  \label{sm-s-eff-rel}
  \delta S(\Delta) \equiv S_{\Eff}(\Delta) - S_{\Eff}(0) = \sum_{\mathclap{\omega_m,\omega_n}}
  \frac{\Delta_{\omega_m}\Delta_{\omega_n}}{16g^2\beta}
  \left[\frac{1}{K}\right]_{\omega_m - \omega_n} \mathclap{-}
  \sum_{\omega_n>0} \log\left( 1 - 
    \frac{\Delta_{\omega_n}\Delta_{-\omega_n}}{4\omega_n^2}\right)\ .
\end{gather}
For any even-f configuration $\Delta_{-\omega_n} = \Delta_{\omega_n}$,
$\delta S(\Delta) \ge 0$, because the first term is positive definite,
and in the second term, $\log(1-\Delta_{\omega_n}^2/4\omega_n^2) < 0$.
Thus without solving the gap equation, one can already conclude that
any even-f state will be unstable.

In the diagonal limit $K_{\nu_n} = \delta_{\nu_n,0} K_0$, the odd-f
solution to Eq.~\ref{sm-gap-eq-2} is
\begin{gather}
  \label{sm-odd-f-sol}
  \Delta_{\omega_n}^- = -\Delta_{-\omega_n}^- = 2 \sqrt{g^2 T K_0 -
    \omega_n^2}\ .
\end{gather}
From Eq.~\ref{sm-s-eff-rel}, the action contributed by
$\Delta_{\omega_n}$ and $\Delta_{-\omega_n}$ is
\begin{gather}
  \delta S_{\omega_n} = 1 - \frac{\omega_n^2}{2g^2 T K_0} - \log \frac{2g^2 T K_0}{\omega_n^2}\ .
\end{gather}
One can verify that $1-x+\log x \le 0$, with equality occuring at
$x=1$, thus the odd-f solution Eq.~\ref{sm-odd-f-sol} has a lower free
energy than the normal state and is stable.

To understand in a more general way why only the odd-f state is
stable, one may adopt the Ginzburg-Landau perspective and expand the
$\log$ term in Eq.~\ref{sm-s-eff-rel} to quartic order in $\Delta$, keeping both, odd-f and even-f components, $\Delta^{\pm}_{\omega_n} = \frac{1}{2}(\Delta_{\omega_n} \pm \Delta_{-\omega_n})$, 
\begin{gather}
  \delta S(\Delta) = \sum_{\mathclap{\omega_m,\omega_n}}
  \frac{\Delta^{-}_{\omega_m}\Delta^{-}_{\omega_n}+
\Delta^{+}_{\omega_m}\Delta^{+}_{\omega_n}}{16g^2\beta}
  \left[\frac{1}{K}\right]_{\omega_m - \omega_n} %
  + \sum_{\omega_n > 0} \left[
    \frac{(\Delta^{+}_{\omega_n})^2-(\Delta^{-}_{\omega_n})^2}{4\omega_n^2} +
    \frac{(\Delta^{+}_{\omega_n})^4+(\Delta^{-}_{\omega_n})^4
-2(\Delta^{+}_{\omega_n})^2(\Delta^{-}_{\omega_n})^2}{32\omega_n^{\,4}}
  \right] + \mathcal{O}(\Delta^6) \ .
\end{gather}
The first term on the r.h.s.~does not have a mixed term owing to the even symmetry of $1/K$. We now immediately see that the odd-f component contributes a negative contribution in quadratic order, which may induce a transition, while the even-f contribution does not. However, while the fourth order terms of both pure odd-f and pure even-f type are positive, the mixed term is negative. This opens the possibility that in an odd-f paired state, even-f pairing may be induced at some lower temperature. In order to explore this scenario we first simplify the analysis by defining frequency independent dimensionless order parameters $A,B$ , where $A$ is the initial slope of $\Delta^{-}$, $\Delta^{-}_{\omega_n}=A\omega_n$ and $B=\Delta^{+}_{\omega_n}/\Omega$ . We restrict the frequency integrations to low frequency, $|\omega_n|<\omega_x$, where the cutoff $\omega_x$ is to be suitably defined later. The above action may then be expressed as a Ginzburg-Landau free energy expansion in the order parameters $A,B$,  
\begin{gather}
F(A,B) \equiv \delta S(\Delta) = c_0 \left[\left(t-1\right) A^2+ \frac{A^4}{8}\right]
+\Bigl[\left(b_0 t + 4c_{1}\right)B^2+ 2 c_{2} B^4\Bigr] -  c_{1} A^2 B^2 + \mathcal{O}(A^6,B^6,\cdots) \ .
\end{gather}
Here we defined $t=\frac{T}{T_c}$ and the positive dimensionless coefficients
\begin{gather}
b_{\ell} = T_c \sideset{}{'}\sum_{\omega_m,\omega_n} \frac{(\omega_m\omega_n)^{\ell}}{16g^2 \Omega^{2(\ell - 1)}} \left[ \frac{1}{K}\right]_{\omega_m - \omega_n} \quad , \quad c_{\ell} = \frac{1}{4} \sideset{}{'}\sum_{\omega_n > 0} \left[\frac{\Omega}{2\omega_n}\right]^{2\ell}\ ,
\end{gather}
the primed sum imposes the aforementioned frequency cutoff $\omega_x$ implicitly determined by requiring
\begin{gather}
  b_1 \stackrel{!}{=} c_0\ .
\end{gather}
If we now substitute the meanfield solution $A^2=4(1-T/T_c)$ for odd-f pairing (assuming $g>g_c$) into the mixed term we find that the determining equation for $B$,
\begin{gather}
  \frac{\partial F}{\partial (B^2)} = (b_0 + 4c_1) t + 4c_2B^2 \stackrel{!}{=} 0\ ,
\end{gather}
does not have a real solution for any positive $t$.

\end{document}